\documentclass{achemso}

\usepackage{amsmath}
\usepackage{gensymb}
\usepackage{color}
\usepackage{hyperref}
\usepackage{natbib}

\newcommand{\johann}[1]{ \textcolor{black}{#1} }

\author{Hadi Arjmandi-Tash}\affiliation{Univ. Grenoble Alpes, CNRS, Institut Neel, F-38000 Grenoble, France}\altaffiliation{Present address : Leiden University, Leiden (The Netherlands)}
\author{Dipankar Kalita}\affiliation{Univ. Grenoble Alpes, CNRS, Institut Neel, F-38000 Grenoble, France}
\author{Zheng Han}\affiliation{Univ. Grenoble Alpes, CNRS, Institut Neel, F-38000 Grenoble, France}
\author{Riadh Othmen}\affiliation{Univ. Grenoble Alpes, CNRS, Institut Neel, F-38000 Grenoble, France}
\author{Cecile Berne}\affiliation{Univ. Grenoble Alpes, CNRS, Institut Neel, F-38000 Grenoble, France}
\author{John Landers}\affiliation{Univ. Grenoble Alpes, CNRS, Institut Neel, F-38000 Grenoble, France}
\author{Kenji Watanabe}\affiliation{Advanced Materials Laboratory, National Institute for Materials Science, 1-1 Namiki, Tsukuba, 305-0044, Japan.}
\author{Takashi Taniguchi}\affiliation{Advanced Materials Laboratory, National Institute for Materials Science, 1-1 Namiki, Tsukuba, 305-0044, Japan.}
\author{Laetitia Marty}\affiliation{Univ. Grenoble Alpes, CNRS, Institut Neel, F-38000 Grenoble, France}
\author{Johann Coraux}\affiliation{Univ. Grenoble Alpes, CNRS, Institut Neel, F-38000 Grenoble, France}
\author{Nedjma Bendiab}\affiliation{Univ. Grenoble Alpes, CNRS, Institut Neel, F-38000 Grenoble, France}
\author{Vincent Bouchiat}\affiliation{Univ. Grenoble Alpes, CNRS, Institut Neel, F-38000 Grenoble, France}
\email{vincent.bouchiat@neel.cnrs.fr}

\title[An \textsf{achemso} demo]
  {High-Yield Proximity-Induced Chemical Vapor Deposition of Graphene Over Millimeter-Sized Hexagonal Boron Nitride}


\begin{document}



\linespread{2}
\begin{abstract}
We present a transfer-free preparation method for graphene on hexagonal boron nitride (h-BN) crystals by chemical vapor deposition of graphene\johann{via a catalytic proximity effect, \textit{i.e.} activated by a Cu catalyst close-by}. We demonstrate the full coverage by monolayer graphene of half-millimeter-sized hexagonal boron nitride crystals \johann{exfoliated on a copper foil prior to growth. We demonstrate that the proximity of the copper catalyst ensures high yield with the growth rate estimated between of $2\,\mu m/min$ to $5\,\mu m/min$}. Optical and electron microscopies together with confocal micro-Raman mapping confirm that graphene covers the top surface of h-BN crystals that we attribute to be a lateral growth from the supporting catalytic copper substrate. Structural and electron transport characterization of the \textit{in-situ} grown graphene present an electronic mobility of about $20,000\,cm^2/(V.s)$. Comparison with graphene/h-BN stacks obtained by manual transferring of similar CVD graphene onto h-BN, confirms the better neutrality reached by the self-assembled structures.
\end{abstract}

\openup 1em

\section{Introduction}
Hexagonal boron nitride has been experimentally identified as an outstanding dielectric material for supporting graphene\,\cite{Dean2010}. h-BN is a large bandgap semiconductor and its 2D lattice shares the same symmetry and almost identical lattice constant with graphene. Furthermore, it possesses a smooth and charge-neutral surface\,\cite{Wang2011}. For all these reasons, h-BN buffer layers have shown to preserve the exceptional electronic properties of graphene\,\cite{Dean2010}, leading to long-range ballistic transport\cite{Mayorov2011,Wang2013a} when graphene is fully encapsulated in between h-BN flakes. Moreover, the environmental 2D superpotential\,\cite{Xue2011} induced on graphene by h-BN (leading to the so-called moir\'{e} pattern) has brought new physics within reach, such as the discovery of fractal quantum Hall effect characterized by an energy spectrum obeying a Hofstater's butterfly pattern\,\cite{Yankowitz2012,Dean2013,Hunt2013}. 

In all these reports, graphene/h-BN heterostructures were prepared by direct transfer of flakes using processes relying on physical adhesion via van der Waals interaction. This implies that graphene (exfoliated or chemically grown) is firstly isolated and then transferred onto h-BN host flakes. This manual critical step involves either micromanipulation of small flakes\,\cite{Dean2010} or transfer of large area CVD-grown graphene sheets based either on wet\,\cite{Xue2011} or dry\,\cite{Petrone2012,Banszerus2015} methods. Wrinkles, trapped air blisters, and polymeric residues are frequently found at the interfaces\,\cite{Haigh2012}. Besides, the multi-transfer procedure has low yield, hence being incompatible with mass-production. Alternative preparation techniques however do exist, and have been reported as early as in 2000\cite{Oshima2000, kawasaki2002}. In these seminal works, graphene was grown by chemical vapor deposition (CVD) under ultra-high vacuum onto a single layer of h-BN prepared on a metal surface. The graphene/h-BN stacks formed accordingly leave the properties of graphene mostly unaffected\,\cite{Roth2013}. The concept has recently been extended to less stringent conditions\,\cite{Wang2013}, and in the absence of a metallic substrate, which is advantageous in the view of electronic transport devices\,\cite{Son2011,Ding2011,Tang2012}. In all these reports however, the yield for graphene growth is quite low (growth duration ranging from several tens of minutes\,\cite{Garcia2012} up to several hours\,\cite{Tang2012}\cite{Yang2013a}), and the mechanisms driving the growth are ill-understood.

Indeed, the h-BN surface is inert towards the decomposition of hydrocarbon at the used growth temperatures. This issue could be circumvented by making the carbon precursor active prior to its adsorption on the surface, in plasma-enhanced CVD\cite{Yang2013}. An elegant and easily implementable alternative was proposed by Kim \textit{et al.}\cite{Kim22013}, who demonstrated the lateral growth of graphene islands nucleating on a copper support and extending on top of single-layer h-BN. Our work elaborates on this approach and establishes a high-yield lateral growth that can also occur on much thicker (up to 1-micron-thick) and on larger (at least above few tenths of a millimeter) h-BN crystals. Our technique avoids the use of hazardous precursors needed for the chemical growth of h-BN, and instead employs large exfoliated h-BN flakes of high crystalline qualities. Note that our h-BN supports are more rigid and flatter compared to few mono-layered h-BN thin films, which make the produced stacks better-suited for easy transfer and handling of heterostructures after the growth, together with electrical shielding of the graphene layer. Indeed, we report a low residual charge carrier density of 4.5$\times$10$^{10}$\,cm$^{-2}$, and an electronic mobility as high as 20,000\,cm$^2$/(Vs) at 80\,K, close to the conductance saturation point. Finally, consistent with Raman spectroscopy and charge carrier density-dependent electronic transport data, we provide a microscopic model describing the scattering of charge carriers in graphene on h-BN.


\section{Direct growth of graphene on h-BN flakes by proximity-catalytic growth}
\begin{figure}
\includegraphics[width=120mm]{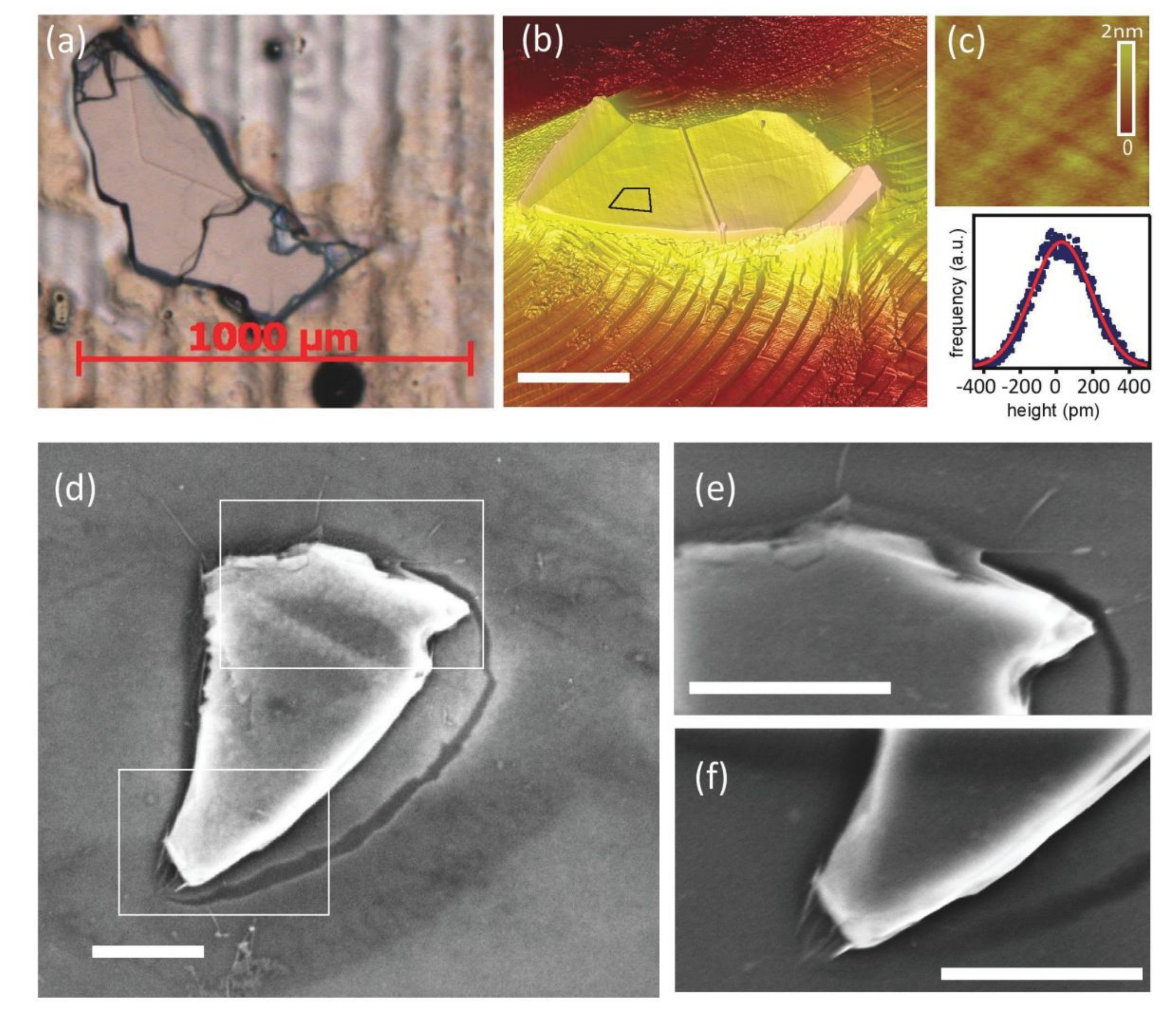}
\caption{(a) Optical micrograph of a large hexagonal boron nitride crystal exfoliated on a copper foil, fully covered with graphene, after the growth. (b) Atomic force micrograph (AFM) of graphene covered h-BN flakes surrounded by the copper foil which exhibits the typical vicinal surface reconstruction. The scale bar corresponds to 2\,$\mu$m (c) High resolution AFM mapping of the mesa (0.4$\times$0.4 $\mu m^2$ area), marked by the overlayed window in (b). The bottom inset shows the histogram of the height distribution corresponding to this mapping. Solid line is Gaussian fits to the distribution. (d-f) Scanning electron micrographs of a h-BN and graphene layer after being transferred on oxidized silicon, showing the continuous graphene veil covering both h-BN flakes together with the host substrate. (e) and (f) are respective zooms on the top and bottom windows marked in (d). Graphene is ruptured around the flakes due to stress during the transfer. Scale bars in d, e, and f corresponds to 500\,nm.}
\label{fig:fig_1}
\end{figure}

The technique, which we refers to as proximity-catalytic growth, is based on mechanically exfoliating large h-BN flakes of areas up to half-millimeter-square and thicknesses ranging from $\sim0.01\,\mu m$ to $\sim1\,\mu m$ onto standard copper foils prior the growth (\autoref{fig:fig_1}-a, see methods for details). The graphene is then grown using either classical\,\cite{Li2009} or pulsed\,\cite{Han2014} CVD techniques, forming a continuous layer with no multilayer patches, in the latter case. For further analysis and device fabrication, graphene/h-BN stacks are then transferred onto oxidized silicon substrates in a dry transfer method. 

An atomic force microscopy (AFM) image of a h-BN flake is shown in \autoref{fig:fig_1}-b. The flake appears to be half-sunk into the copper substrate, as a result of the semi-liquid surface of the copper during the growth. The top surface of the h-BN flake keeps a very low roughness, as  demonstrated by the high resolution AFM micrograph (\autoref{fig:fig_1}-c) of the top layer taken from an area defined by the overlayed window in \autoref{fig:fig_1}-b. A histogram of the height distribution measured inside this window (bottom inset) reveals a roughness of $\sim$ 4\,\AA, which is consistent with the reported roughnesses of h-BN and grphene on h-BN\cite{Dean2010}. We can then conclude that the surface quality of the h-BN is preserved during the high temperature process. \autoref{fig:fig_1}-d to f show scanning electron microscopy (SEM) images of a graphene/h-BN stack, after being transferred on an oxidized silicon wafer. It is worth noting that continuity of the graphene top veil is observed as the graphene grown on h-BN and on surrounding are still connected. Ripples and bridges of locally suspended graphene are indeed observed at some of the edges of the h-BN flake. 

\begin{figure}
\includegraphics[width=160mm]{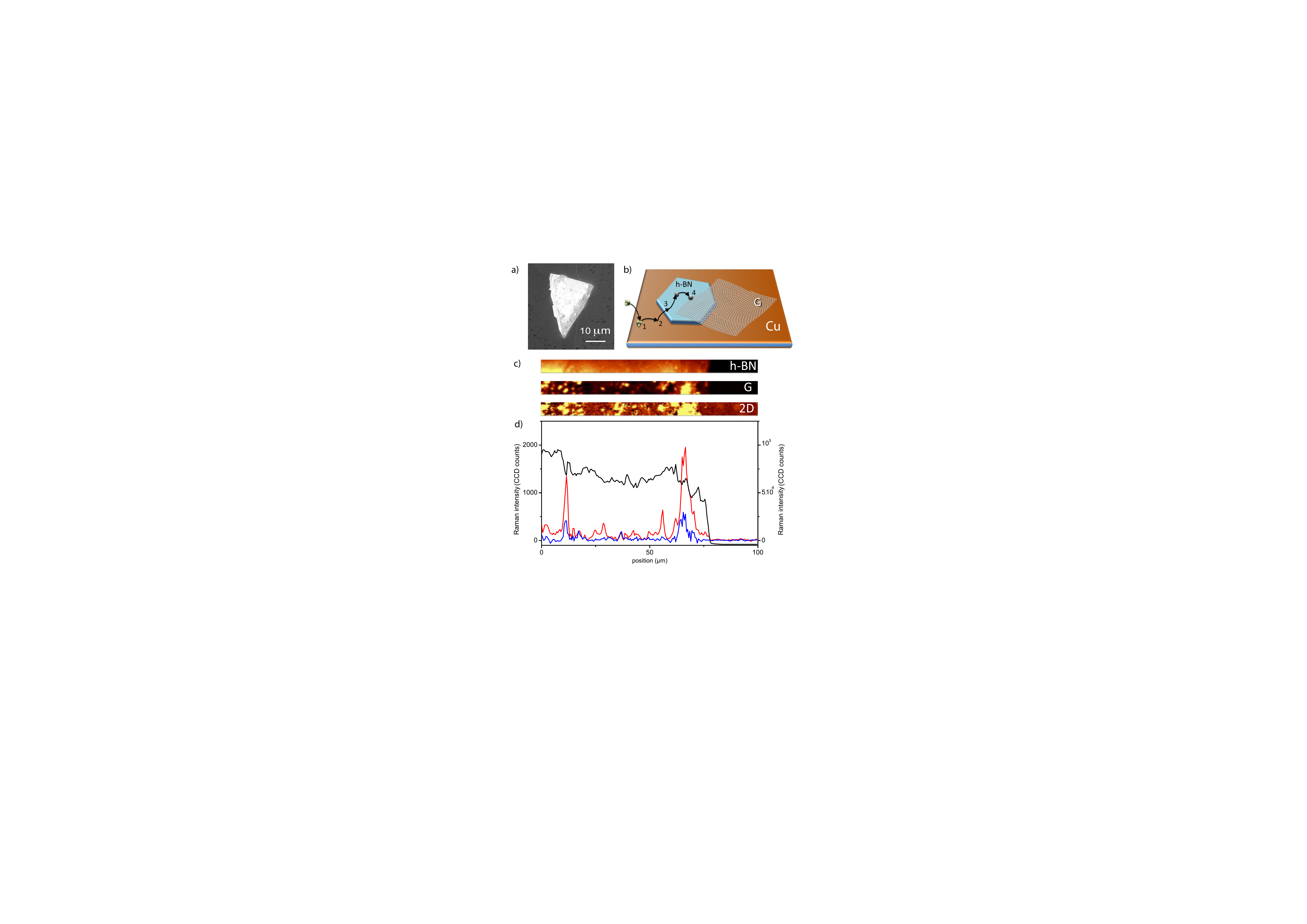}
\caption{\textcolor{black}{a) SEM micrograph of a h-BN flake on Cu after 90\,s growth time, showing h-BN flake partially covered with graphene on a distance of $\sim\,4\,\mu m$ to $\sim\,8\,\mu m$ from the edges. b) Mechanism of CVD proximity growth, adapted from Ref.\,\cite{Kim2013b}. The carbon precursor is cracked on the Cu catalyst surface (step 1), the resulting carbon adatoms randomly diffuse on Cu foil (step 2) and climb up the h-BN (step 3) to eventually assemble on the $sp^2$ lattice (step 4). This last step has been identified as the time-limiting step, therefore leading to similar kinetics for graphene growth on Cu and on h-BN. c) Raman mappings recorded at the edge of the same h-BN flake as in a. d) Averaged Raman line scans of the mappings in c; black: h-BN (corresponding to vertical axis on right), Blue: G-mode and red: 2D mode (both corresponding to the vertical axis on left). }}
\label{fig:fig_new}
\end{figure}

\textcolor{black}{ 
\section{Growth mechanism of graphene onto exfoliated h-BN flakes}
}
\textcolor{black}{
h-BN is known to be a chemically inert material well above 1000$^\circ$C. Furthermore, no substantial decomposition of methane should occur on h-BN, and a methane molecule should rapidly desorb without a chance of reacting with other molecules to form graphene. Several reports however revealed that the neighbouring Cu, known to catalytically decompose methane (this is the essence of graphene CVD on bare Cu), generates the active C species for the growth of graphene away from the catalytic surface itself. Such a scenario was invoked to account for the growth of graphene on single layer h-BN on Cu\,\cite{Kim2013}. We note that the decomposition process may also occur already in the gas phase, inside the reactor, due to the presence of Cu, \textit{e.g.} on the walls of the reactor\,\cite{Kim2013}. However, if at play, the later effect should lead to the formation of graphene not only on h-BN, but also on the graphene-covered Cu, which we do not observe. Another possibility is that the h-BN surface is turned catalytically active by Cu in provenance from the surrounding Cu foil. At our growth temperatures, Cu cannot wet h-BN, but rather would form clusters. The presence of such clusters is ruled out by our AFM measurements, showing a flat surface (\autoref{fig:fig_1}-c). At this point we can describe the main steps of the growth of graphene on h-BN as follows (\autoref{fig:fig_new}-b): Methane is adsorbed at the Cu surface where it is readily decomposed into carbon adatoms. On the contrary, when landing on h-BN, methane is rapidly desorbed. As it is well known on Cu\,\cite{Kim2012} and other low-C-solubility metals\,\cite{Coraux2009}, the limiting step during graphene growth is the incorporation (attachment) of C at the graphene edges, while the diffusion of C adatoms is a comparatively very fast process. The later implies that the nucleation is heterogeneous, occurring at defects of Cu, while very improbable at the h-BN surface where -- as it is the case on a graphene surface as well\,\cite{Lehtinen2003} -- diffusion is considerably faster. After the nucleation stage, graphene grows, finding in its surrounding a sufficient concentration of carbon adatoms to do so. Noteworthy, carbon efficiently diffuses all around the surface, including on Cu, on h-BN as we just discussed, but also across the frontier between Cu and h-BN, even in the case of flake thickness as high as a few micrometers. We find that within 20\,min, graphene growth fronts typically cover at least 100\,$\mu$m (in \autoref{fig:fig_2}-d-g, a large h-BN flake is fully covered with graphene); When stopping the growth after 90\,s , the complementary SEM and Raman analyses (\autoref{fig:fig_new}-a,c, and d) reveal that the borders of h-BN flakes are covered by graphene on a distance of $\sim4\,\mu m$ to $\sim8\,\mu m$, which provides an estimation of $2\,\mu m/min$ to $5\,\mu m/min$ for the growth rate, a larger value than in recent reports of growth on h-BN describing the use of a silicon carbide substrate\,\cite{Mishra2016} or of another deposition using molecular beam epitaxy\,\cite{Garcia2012}\cite{Garcia2013}. Note however that such a growth speed is in agreement with values reported for graphene growth on pure Cu\,\cite{Wu2016}, further strengthening the scenario of a proximity growth with carbon adatoms emerging from nearby Cu. 
}

\section{Raman characterization of synthesized graphene}
 
\begin{figure}
\includegraphics[width=160mm]{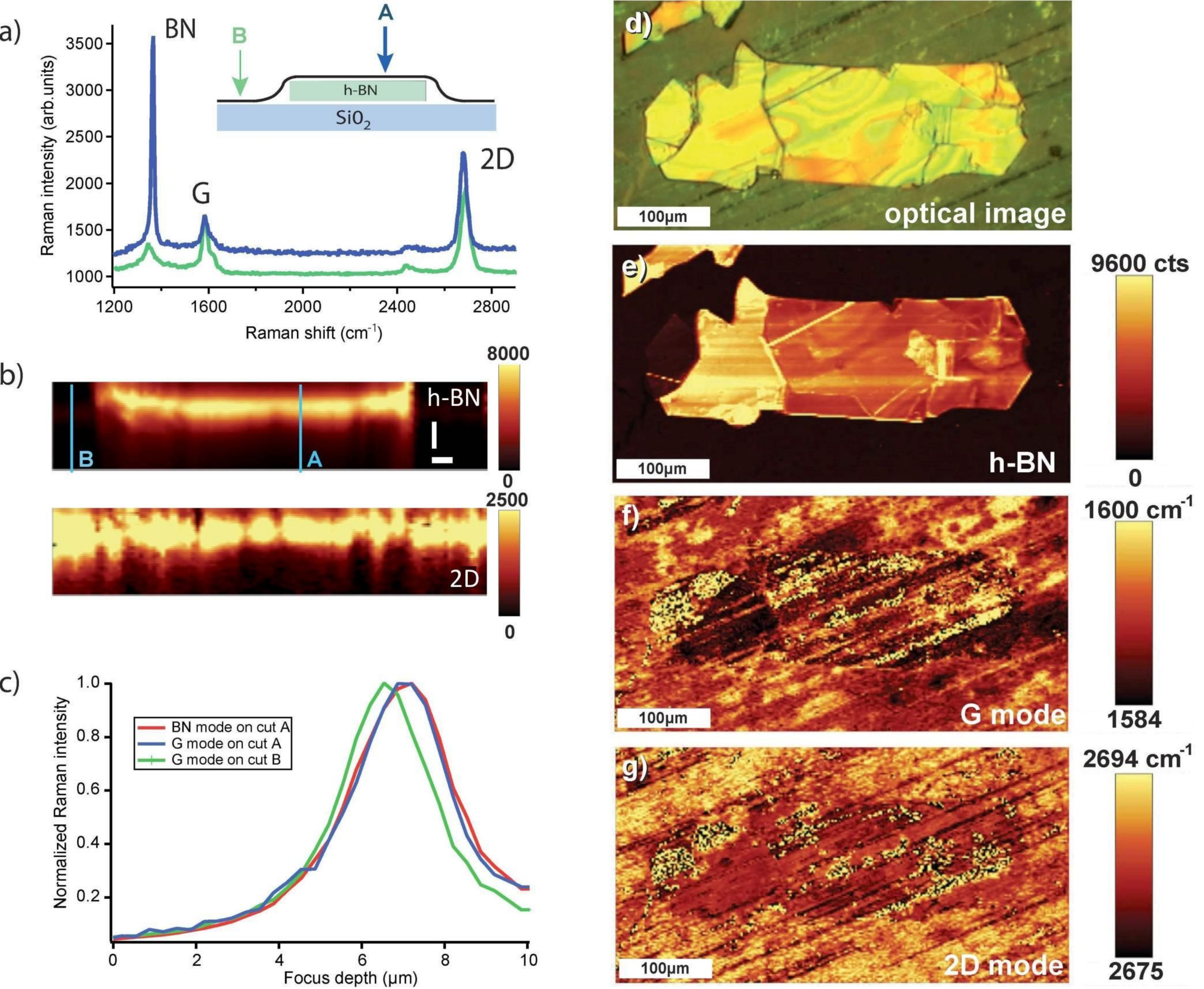}
\caption{Raman Analysis of a direct grown Graphene/Boron Nitride stack after transfer onto an oxidized silicon substrate. a) Raman spectra obtained at 532nm are recorded on a h-BN flake (arrow A, blue) and on SiO$_2$ (arrow B, green)  b) shows Raman maps at 532 nm obtained by recording the Raman spectra at different vertical focus planes. They provide vertical cuts of the Raman intensity sum in the [1344-1377 cm\textsuperscript{-1}] frequency range (top: h-BN mode) and in the [2652-2706 cm\textsuperscript{-1}] frequency range (bottom: 2D mode of graphene). Horizontal scale bar is $10\mu$m, vertical one is $3\mu$m. c) Cuts of maps b) obtained along the lines A and B. The graph shows the vertical evolution of h-BN and G modes. (d) Optical microscope image of a direct grown Graphene/Boron Nitride stack  transferred onto an oxidized silicon substrate. (e-g) Raman maps of the same flake showing: e) integrated Raman intensity in the [1344-1377 cm\textsuperscript{-1}] frequency range (h-BN mode); f) frequency of the Lorentzian fit in the [1500-1615 cm\textsuperscript{-1}] frequency range (G mode of graphene); g) frequency of the Lorentzian fit in the [2600-2800 cm\textsuperscript{-1}] frequency range (2D mode of graphene).}
\label{fig:fig_2}
\end{figure}

We performed confocal Raman analysis and microRaman spectroscopy maps on many graphene/h-BN flakes of various sizes, before and after transfer from the copper foil. \autoref{fig:fig_2}-a reveals that the typical Raman signature\cite{Ferrari2007} of monolayer graphene (G and 2D bands) is found everywhere on the sample including at the position where h-BN crystals are present. At these positions, the characteristic $E_\mathrm{2g}$ peak of h-BN is additionally detected (\autoref{fig:fig_2}-a, blue curve). The overlapping of graphene and h-BN signatures further confirm the overgrowth of graphene on h-BN. \autoref{fig:fig_2}-b show depth (y-z) mapping of the intensity of the $E_\mathrm{2g}$ mode of h-BN and of the 2D mode of graphene across a line corresponding to the right side of \autoref{fig:fig_2}-d. Cutting profiles via the marked lines A (through h-BN) and B (outside h-BN) are also plotted in \autoref{fig:fig_2}-c. Note that the apparent thickness of the layers is convoluted by the beam waist of the optical setup which is about 2 microns. Interestingly we observe a shift in the position of the maximum intensity of the 2D peak measured through the lines: the shift matches the thickness of the h-BN flake and is attributed to leveling graphene up by h-BN due to the top coverage of the h-BN with graphene. 

\autoref{fig:fig_2}-f and g analyze the position of the G and 2D modes of graphene on the whole area. No remarkable discrepancy can be identified on h-BN and on copper foil and the frequencies are always in expected ranges. Raman analyses clearly confirm, \textcolor{black}{ down to a 300\,nm resolution,} that graphene evenly covers large h-BN crystals, with no apparent limitation of coverage size (at least up to half a millimeter).

\begin{figure}
\includegraphics[width=120mm]{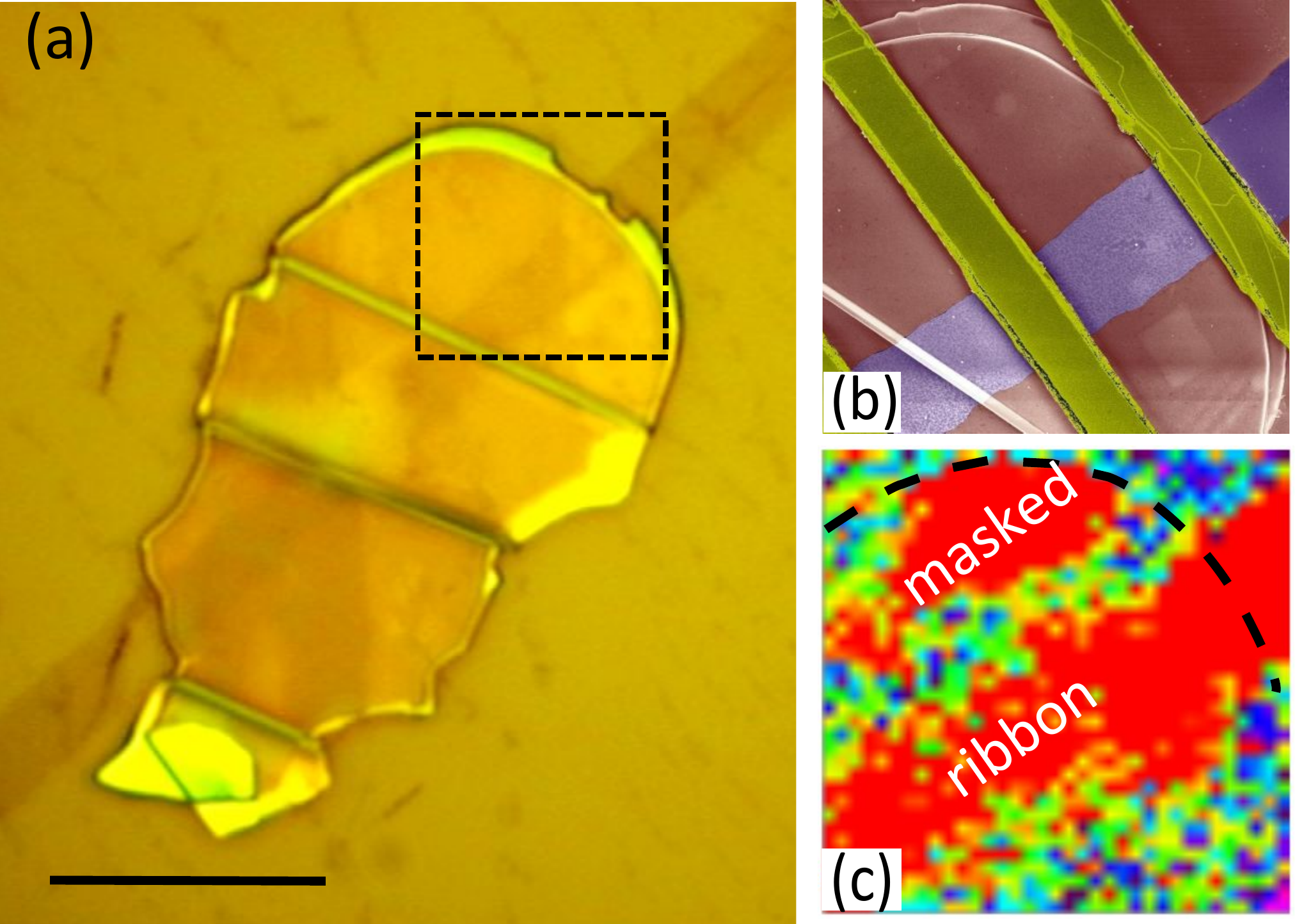}
\caption{Analysis after transferring graphene/h-BN stacks on a silicon wafer:\\ 
a) Optical micrograph image of a  h-BN flake. The scale bar here measures $10\,\mu m$. b) False color SEM image of the area defined by the dashed lines in (a), and (c) the corresponding intensity map of the Raman G-band of the graphene. The border of the h-BN flake is identified by the dashed line here. Graphene ribbon and area masked by the flake are marked in this map.}
\label{fig:Fig5new}
\end{figure}

We continue the discussion by characterizing another graphene/h-BN stack (growth duration: 90\,s) after transferring onto an oxidized silicon wafer. Oxygen etching after transfer to SiO$_2$, is used to pattern graphene into a ribbon. The width of the resulting ribbon (\autoref{fig:Fig5new}-a-c), is identical on top of h-BN and away from it, which would not be possible if graphene lied between h-BN and copper (before transferring) or between h-BN and SiO$_2$ (after transferring). \autoref{fig:Fig5new}-c also reveals the presence of another graphene region which shares a common edge with the h-BN flake indicating that it has been masked by the h-BN flake through the plasma process. This region is not detected in the optical (\autoref{fig:Fig5new}-a) nor the SEM (\autoref{fig:Fig5new}-b) images and thus is not lying on the surface, hence on the backside. We note that Raman visibility of the graphene through covering h-BN flakes is a known phenomenon\,\cite{Wang2012a}. The observation of the growth of graphene at the backside of h-BN is significant as it qualifies the technique for \textit{in-situ} growth of graphene in fully encapsulated systems.

\section{Electron transport properties of graphene/h-BN devices}
\begin{figure}
\centering
\includegraphics[width=1\textwidth]{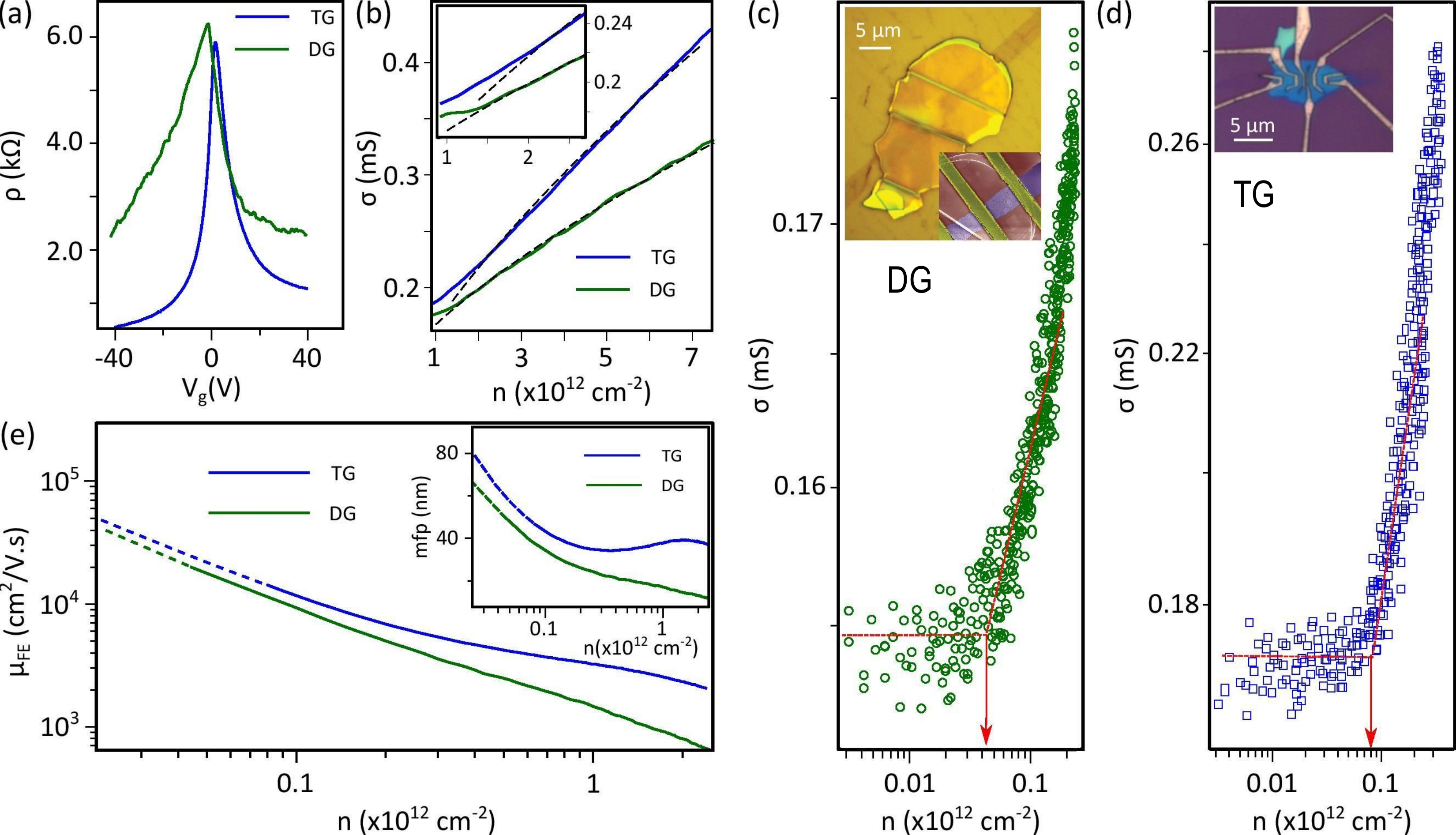} 
\centering
\caption{Comparison of the transport properties of the proximity grown graphene on h-BN (\textit{DG}) with a similar device made by wet transferring CVD graphene (same growth) on h-BN (\textit{TG})\\
a) Gate dependence of the resistivity of the samples, b) Corresponding conductivity of the samples on the electron side: Dashed lines are fits to the curves using the \textit {midgap states} theory described in the main text. Insets to this figure are the semi-log plot  showing low charge density regime. The vertical and horizontal axes have the same units as the main panel. c) and (d) Logarithmic-scale plot of the conductivity of DG and TG samples versus carrier densities close to the Dirac point: The arrows show critical densities below which the conductivity of the samples saturates due to the formation of the electron-hole puddles. Insets show the optical micro-graph of the corresponding samples. e) Comparison of the field effect mobilities $(\mu_{FE})$ and mean free paths (mfp, inset figure) of the charge carriers corresponding to the \textit{in-situ} grown and transferred graphene/h-BN stacks\\
All the measurements have been done at 80\,K.
}
\label{fig:fig 5}
\end{figure}

A lithography pattern followed by oxygen plasma etching enables to remove most of the graphene and to prepare a graphene ribbon that is crossing a h-BN crystal. By depositing electrodes on the ribbon, we fabricate a connected device using this graphene sample, directly grown (noted \textit{DG}) on h-BN. For the sake of comparison, we will compare its electronic properties with a transferred graphene (noted \textit{TG}) sample obtained from a similarly grown graphene sample but via wet transfer onto a h-BN flake, which has been exfoliated and stamped  directly on the host substrate. All the electron transport measurements were performed at 80\,K and are summarized in \autoref{fig:fig 5}.
\\
\autoref{fig:fig 5}-a shows the resistance as a function of the back-gate voltage of the two samples. The charge neutrality points for both samples are located very close to zero voltage which is an indication of a clean graphene on a neutral h-BN substrate. The sample TG shows a sharper peak which is a signature of high carriers mobility. The curve is quite asymmetric in the electron and hole sides which could be due to the electrode charging effects\,\cite{Wu2012}. The asymmetry for the DG sample is more pronounced; process residuals on the sample could be a reason for that. While the residual resistance (resistance at the gate voltages very far from the Dirac point) is quite high for TG sample, it is even higher for the DG one; this observation together with the presence of the resistivity fluctuations at high gate voltages points out that both of the samples and especially the DG one suffer from the high population of the short range crystalline defects. In CVD graphene such defects can happen during the growth, transferring or fabrication process. Since both the samples have passed similar fabrication steps, this observation in the DG sample can be attributed to the formation of the defects during the growth. 

There are few models to describe the scattering near the Dirac point in graphene\,\cite{Hwang2007}\cite{Adam2007}. However since the presence of a D peak in the Raman analysis suggested the existence of the crystalline vacancies in our sample, we use a model which considers scattering by the \textit{mid-gap states}\,\cite{Stauber2007}. In this model, strong disorder associated with the voids are modeled as deep potential wells\,\cite{Stauber2007}\cite{Hentschel2007}:
\begin{align*}
\sigma = \frac{2e^2}{h} \frac{k_F^2}{\pi n_d}(\text{ln}k_FR_0)^2.
\end{align*}

Here, while $k_F$ is the Fermi wave vector of graphene, $n_d$ and $R_0$ refer to the density and characteristic size of the defects, respectively. Black dashed lines in\autoref{fig:fig 5}-b are the fittings with this model, successful to describe our results to a large extent. Table \ref{tbl: mid-gap_states} summarizes the results of this fitting. For the sake of comparison, expected values for an exfoliated graphene\,\cite{Stauber2007} on typical substrates also are included in this table:
\begin{center}
\begin{table}[H]
\captionof{table}{Defect parameters for different graphene samples}
\centering
\begin{tabular}{ccc}
\vspace{5pt}
& $n_d\,\left[cm^{-2}\right]$ & $R_0$\,[\AA]\\ 
\hline 
\\
\vspace{5pt}
present work, TG & $2.7\times 10^{12}$ & $1.3$ \\
\vspace{5pt}
present work, DG & $4\times 10^{12}$ & $1.5$ \\
\vspace{5pt}
exfoliated graphene\,\cite{Stauber2007} & $<10^{11}$ & $1.4$ \\
\hline
\end{tabular}
\label{tbl: mid-gap_states}
\end{table}
\end{center}
\vspace{-2cm}
Clearly the size of the defects in our devices is in the atomic range and matches the predictions. However the population of such defects in our samples are at least one order of magnitude higher than the prediction for pristine graphene, which tends to suppress the mobility of charge carriers in both of the samples. Comparison of the data for TG and DG devices reveals that the indirect access to the catalyst during the growth of the DG sample leads to increasing the population of defects, however the amount of defects added in proximity-driven growth compared to the normal CVD process is much lower than the amount that a CVD grown graphene bears comparing to the predictions.

We note that, unlike the TG, the graphene of the DG sample was sandwiched and protected between a PMMA layer and h-BN flake during the transfer, thus extra vacancies have happened during the growth. Assuming that these vacancies are homogeneously spread all over the graphene and by considering the devices geometries, the average spacing between the voids is estimated to be around 8\,nm. We will see later (\autoref{fig:fig 5}-e) that at high carrier densities (when the role of the defects are important), the mean free path of the electrons in this sample falls below 10\,nm which is very comparable to the spacing between the defects i.e. the transport of the charge carriers are affected by the formation of such voids. 

The inset \autoref{fig:fig 5}-b focuses on the vicinity of the Dirac point. Here we see that by reducing the density of electrons, the model fails to follow the experimental results anymore. Around this area, the charged impurities close to the graphene sheet make a random network of 2D electron and hole puddles which affects the conductivity of the samples\,\cite{Martin2007}. Interestingly, this transition happens at a higher carrier density for the TG sample. By plotting the low density regime in a logarithmic-scale (\autoref{fig:fig 5}-c and d), one can see by approaching the Dirac point, the conductivity of both of the samples reduces monotonically but saturates after a threshold density ($n_{sat}$, depicted by arrows there); The saturated conductivity for the samples ($\sigma_{sat\, DG}\approx 1.0\times\frac{4e^2}{h}$ and $\sigma_{sat\,TG}\approx 1.1\times\frac{4e^2}{h}$) are very close to the universal minimum conductivity predicted for the graphene\,\cite{Stauber2007} which happens below the threshold densities of $n_{sat\,TG}\approx 8\times 10^{10} cm^{-2}$ and $n_{sat\,DG} \approx 4.5\times 10^{10} cm^{-2}$ respectively. The ratio of the saturation density and the corresponding conductivity is proportional to the population of the charged impurities\,\cite{Adam2007}: $n_{imp} \propto (\frac{n_{sat}}{\sigma_{sat}})$; this implies $n_{imp\,DG}/n_{imp\,TG}\approx 0.6$. This number shows the advantage of the utilized direct growth of the graphene in minimizing the charged impurities. The thicknesses of the DG flake is around 80\,nm and knowing that the impurities located with a distance more than $\approx 10\,nm$ from the graphene have a tiny effect on its conductivity\,\cite{Chen2009a}, the estimated impurities in DG sample are either located on top of the graphene or are some impurities in the h-BN flake which might have migrated close to the surface during the growth.

Now we use the equation $\mu_{FE} = \sigma/en$ to calculate the electronic mobility of the samples. The results are plotted in \autoref{fig:fig 5}-e. Atomic scale crystalline vacancies can scatter charge carriers in a range comparable with their size (short range scatterers). As a result, the crystalline vacancies are effective scatterers only when the population of the charge carriers are comparable to the population of the defects\,\cite{Hwang2007}. This is in contrast to charged defects which affect the transport of the charge carriers \-- via Coulomb interaction \-- even if they are very far apart e.g. with much lower charge carrier concentration. The effect of the vacancy and charge impurities on the mobility have been analyzed theoretically. Indeed it is confirmed that very far from the Dirac point ($|V_g-V_D|\sim 100\,V$) the presence of the point defects (with similar population as our samples) may suppress the field effect the mobility up to one order of magnitude\,\cite{Stauber2007}. Comparison of the mobility of our samples also confirm the short range nature of the vacancy defects: close to the Dirac point, the difference between the field effect mobility of the devices (for constant \textit{n}) is very negligible, while by increasing the charge density, the difference of the mobility corresponding to the samples with less (TG) and high (DG)  crystalline defects considerably increases.

The calculated field effect mobilities are valid down to the onset of conductivity saturation \,\cite{Du2008}. The maximum mobility of TG at $n_{sat\,TG}$ is about $14,000\,cm^{2}/Vs$, however the mobility of the DG continuously increases and approaches $20,000\,cm^{2}/Vs$ close to its saturation point. This mobility is one of the highest reported so far for graphene directly grown on h-BN in different techniques. Having this curve, the corresponding mean free path (mfp) of the electrons can be calculated using the equation: $l=(h/2e)\mu\sqrt{n/\pi}$, the results of which are shown in the inset of this figure. Like the mobility, early saturation of the conductivity for the DG, accounts for higher \emph{meaningful} mean free paths for the DG. 

\section{Conclusion}

In conclusion, this work introduces a novel modality in the high-yield chemical vapor deposition of graphene on thick non-catalyst materials (h-BN in our case) where a nearby catalyst indirectly promote the growth through collateral proximity effect. Two important consequences emerge from the presence of catalyst in direct vicinity of the flakes: firstly, as the temperature during growth does not affect the surface quality of the h-BN flakes, most recipes developed for copper foils can be readily used for that growth, leading to full coverage of h-BN crystal with little to no change of parameters. Secondly, the presence of the catalyst ensures that a high coverage speed while strictly monolayer growth can be obtained. We show that the approach is capable to achieve full coverage of graphene with good kinetics (up to 5 microns per min. ) covering milimeter-sized exfoliated hexagonal boron nitride flakes in less than 20 minutes. Further Raman analysis confirms the possibility of growing graphene at the backside of the flake or possibly  in between the h-BN layers. The electronic characterization shows that at high charge carrier density regime, atomic-scaled vacancies hampers the electric transport of our graphene. The devices best perform close to the Dirac point, where the electron scattering is limited by minimizing charged impurities at graphene/h-BN interface, delaying the formation of electron-hole puddles. At this regime, we obtained remarkably high carrier mobility which outperforms normally transferred graphene/h-BN samples.

\begin{acknowledgement}
The authors are grateful for the help from NanoFab team of Institut NEEL. H. Arjmandi-Tash acknowledges grant support from the Nanoscience Foundation of Grenoble. This work is partially supported by the French ANR contracts SUPERGRAPH , CLEANGRAPH and DIRACFORMAG,  the EU contract NMP3-SL-2010-246073, ERC, EU GRAPHENE Flagship and the Region Rhone-Alpes CIBLE program.
\end{acknowledgement}



\bibliography{REFCH2}
\end{document}